\documentstyle[12pt,draft]{article}

\input{tcilatex}
\input tcilatex
\begin{document}

\author{S. Stan\thanks{%
Corresponding author (the present address: Department of Food Sciences and
Technology, Oregon State University, Wiegand Hall 240B, Corvallis, OR
97331-6602, USA)} \\
Faculty of Natural Sciences,\\
Center for Malting and Brewing Science,\\
K.U. Leuven, Kardinaal Mercierlaan 92, \\
B-3001 Heverlee, Belgium \and F. Despa \\
Faculty of Natural Sciences,\\
Department of Physics, \\
K. U. Leuven, Celestijnenlaan 200D,\\
B-3001 Heverlee, Belgium}
\title{On the Doublet Formation in the Flocculation Process of the Yeast Cells }
\maketitle

\begin{abstract}
The combination of single cells to form doublets is regarded as the
rate-limiting step of flocculation and requires the presence of surface
proteins in active form. The process of activation of the flocculation
proteins of yeast cells is described in the frame of the autocrine
interaction regime (Cantrell, D. A. and Smith, K. A., 1984, Science 224,
1312-1316). The influence of several effectors (the cell efficiency to use
sugars, the calcium content in the external medium and the probability that
free cells collide each other under thermal motion conditions) on the
initial rate of flocculation and on the fraction of remaining free cells in
the steady state is briefly discussed in the paper. The present model offers
an useful tool for further quantitative investigations in this topic. Also,
it indicates qualitatively a way in which the regulation of flocculation
might be controlled at the level of the expression of cell-surface
activation abilities.


\newpage{}
\end{abstract}

\section{Introduction}

Flocculation is a well-known example of natural, active aggregation and is
defined as the reversible aggregation of cells into flocs (Stradford, 1992;
Straver {\it et al.}, 1993). Particularly, the flocculation process is
important in industrial processes related to fermentation technology
(brewing, wine-making, bioconversions).

In essence, flocculation is an ongoing process in which a yeast cell
population, initially consisting entirely of single cells, is transformed
into two fractions, flocculated and single cells. This results in a bimodal
floc-size distribution (Davis and Hunt, 1986), which means that there are
single cells and flocs with a clear dividing line between them, not a whole
spectrum of sizes of miniflocs spreading themselves over large distances in
the solute volume. Also, the visual examination of single cells and small
flocs falling under gravity showed that large fast-moving flocs did not
collide with single cells in their path but swept them aside in mass flows
of liquid ahead of the particles (Stradford, 1992). The only collision
observed, and subsequent adhesion, were between floc particles of
approximately similar sizes.

In the light of this evidence, Stratford (1992) proposed the cascade theory
to approach the flocculation process. According to this theory, the
particles only succeed in colliding with particles of similar size. The
rate-limiting step of the process is the combination of single cells to form
doublets. Flocs then rapidly build up in size by combining with
similar-sized flocs until they reach the maximum size limit imposed by shear
forces of agitation. As a consequence of rapid formation of large flocs, the
relocation of the remaining single cells and floc compression elapses. The
overall effect is that flocs become progressively less dense and incorporate
more empty space as size increases (Davis and Hunt, 1986). In the empty
spaces of the large floc structure, the remaining single cells form small
clusters. Such a system showing local symmetry repeated at different
enlargements and scales is a fractal structure (Mandelbrot, 1990). The
fractal dimension measured for yeast flocs confirms the supercluster nature
of the floc structure (Davis and Hunt, 1986), and also indicates that the
structure was formed by a diffusion-limited process (Schaefer, 1989). Thus,
the properties of the floc structure are strongly related to their
microstructured morphologies resulting from a specific agglutination and
growth process.

Since major macroscopic features of the flocculation process seem to be well
understood, several questions regarding the microscopic aspects of the
process still remain open. For example, it is of a crucial importance to
know how two cells form a dimer structure? How will factors in the external
cellular medium prevent or induce the dimer formation and which are the main
external factors intervening in the mechanism of dimer formation?

Within a fully developed yeast-culture, most cells are flocculating or
retain the option to flocculate when activated by calcium ions. The
flocculation process of the yeast cells requires the presence of surface
proteins and mannan receptors. If these are not available, or masked,
blocked by binding specific sugars or generally inhibited or denatured,
flocculation can not occur. Flocculation, once developed, is an intrinsic
property of the cell wall. To sum up at this point, sugar-binding proteins,
lectins and flocculation share the characteristics of inhibition by specific
sugars and a requirement for calcium ions. Also, for flocculation to be
expressed, receptor groups must be available on the other yeast walls to
allow bonding by the flocculation protein (Stradford, 1992).

We are entering an exciting phase within which links are being forged
between transduction events at the plasma membrane and the surface cell
receptors (carbohydrate and proteins), which contribute to the onset of
flocculation. Indeed, there are several transduction steps to elucidate in
order to understand how the regulatory factors (antigens) act on the genes
involved in the protein secretion, and how the secreted proteins become
surface cell receptors and bring about the onset of flocculation.
Nevertheless, the precursor stage of dimerization, after the flocculation
proteins are fully expressed in the cell wall but not activated yet for the
flocculation onset, is rather complex. Further investigation at this stage
is one pressing issue in the general problem of understanding how the
flocculation effectors govern the dynamics of the process and how can the
regulation of flocculation can be controlled at the level of the expression
of cell-surface activation abilities.

With this background we can state the main issues of this paper. What is the
precursor stage of the dimer formation and how do we characterize it? What
is the physical mechanism of the process and what mathematical equation
governs this process?

To address these topics the objective of this study was twofold: First, we
aimed to demonstrate that the process of activation of the flocculation
proteins can be seen in a more general context of cellular processes, that
is the autocrine interaction regime (Cantrell and Smith, 1984). In turn,
this would mean flocculation has a kinetic base and informations can at
least be mathematically attainable on a computer (Despa and Despa, 1997). A
second objective was to explore quantitatively the influence of the
flocculation effectors on the rate of the dimer formation. Beyond this was
the idea of varying both the cell intrinsic parameters and external
effectors to see how modifying their range would affect, within the
limitations of the present approach, the tendency of cells to form dimers.

\section{Theoretical model}

In order to achieve a rapid progress in describing flocculence, we will use
in the following a more simplified model which, indeed, retains the
important features observed so far.

Suppose that the conditions for binding between specific surface lectin
proteins of flocculent cells and carbohydrate receptors on nearby cells are
fulfilled satisfied. This is the lectin-like binding hypothesis (Miki et
al., 1982). The surface proteins need active conformations (lectin
properties) in order to bond to the corresponding carbohydrate receptors.
This is fully ensured by bonding $Ca^{2+}$ ions to the flocculation
proteins, which lead consequently to their lectin (active) conformation.
Note that, another hypothesis of the doublet formation assumes the leading
role to the calcium ions (the calcium-bridging hypothesis (Harris, 1959;
Mill, 1964)). There, the divalent calcium ions form bridges between surface
carboxyl or phosphate groups on different cells. The calcium-bridging
hypothesis falls to explain the inhibition of flocculation by sugars while,
the lectin-like binding hypothesis succeeded. The lectin-like binding
hypothesis gets further support by the observation that various
non-flocculent strains of yeast are able to co-flocculate when mixed
together (Stradford, 1992a).

We recall from above that the onset of the flocculation process needs a
primary insertion of the flocculation proteins from the plasma membrane into
the cell wall and is based on the calcium-dependent interaction between
lectins and cell wall mannan receptors. The mechanisms leading to the
activation of the flocculation proteins (this implicitly assumes correct
lectin conformation by $Ca^{2+}$ binding) are largely unknown. In the
following, we propose an activation mechanism for the flocculation proteins
much in the same manner as it has used to describe the activation of helper $%
T$-cells by $IL-2$ growth factors (Cantrell and Smith, 1984; Despa and
Despa, 1997). There, the interaction between the helper $T-$ cells and their
corresponding $IL-2$ growth factors obey a self-interaction mechanism within
the limit of the autocrine binding regime.

Similarly, we assume here that the activation mechanism of the flocculation
proteins and the onset of the flocculation process involve an autocrine
binding regime for sugar radicals and $Ca^{2+}$ ions. It implies a
self-interaction phenomenon, which means the following: The presence of
sugars (and the other nutrients) in the autocrine region leads primarily to
its consumption and consequent production of metabolic energy. Production of
metabolic energy is vital at all stages of the cell development including
the protein encoding by flocculation genes (Novick {\it et al}., 1981). Once
the flocculation proteins developed, these are inserted in the cell wall
and, afterwards, activated by $Ca^{2+}$ binding. Concomitantly, the
activated surface proteins may be blocked to onset flocculation by the
inhibitory action of sugars. Lectins may have great affinity for wort sugars
but the interaction in the wall mannan of the cell with other sugar residues
may be also possible. So that, on one hand the sugar promotes flocculation
by providing the cell with the metabolic energy for flocculation protein
formation and, on the other hand, it fetters the ongoing process by specific
binding to the active flocculation proteins.

In the following, we consider that the surface proteins are preformed at an
earlier stage of development and inserted in the cell wall. The abundance of
flocculation proteins in the cell wall is direct related to the cell
efficiency to use sugars (nutrients). We also assume that the calcium ions
have the physical access to the flocculation proteins. The active
conformation of the flocculation proteins is achieved by bonding $Ca^{2+}$
ions. Accordingly, we propose the following rate equation 
\begin{equation}
\frac{dW}{dt}=\alpha +N_{bound}\left( t\right) -\alpha _{-1}W\left( t\right)
\;\;,  \tag{1}
\end{equation}
to describe the activation of the surface proteins. The first term in the
right side of this equation represents the cell efficiency to use sugars
(nutrients) in order to produce the metabolic energy needed for the
activation of the flocculation proteins. (Actually, the ratio of
concentration of signal nutrient to sugar may conceivably influence
flocculation (Stradford, 1992).) The effective value of this term accounts
for the sugar depletion in unit time. The second term, $N_{bound}\left(
t\right) $, represents the time-depending number of receptors (flocculation
proteins) where the $Ca^{2+}$ ions are bound. This depends on the size of
the autocrine region (i.e., the region close to the cell in which any $%
Ca^{2+}$ ion will be undoubtedly bound to the surface protein). The third
term represents the self-interaction term, which induces a saturation effect
due to the extra-sugar content ($\alpha _{-1}\equiv 1-\alpha ,$ in the
absolute value).

In such a way, we may observe that the external concentration of sugar
controls both early events, prior the activation, (starting the signaling
cascade of proteins encoding by structural genes) and later events (e.g.,
the inhibition of the surface cell receptors). $Ca^{2+}$ has the secondary
role to promote the lectin properties of the surface proteins. Other
cumbersome effects related to the cofactors action on the flocculation
proteins (proteolytic cleavage, physical exposure) are disregarded in the
present model.

Further on, let us consider a suspension of cells at the moment of time $t$,
each of them having an average number $W_{0}$ of surface proteins uniformly
distributed on the cell surface. We assume that the calcium ions having the
concentration $n_{Ca^{2+}}$ move diffusively around the yeast cells. The
surface proteins bind calcium ions from the autocrine region resulting
flocculation proteins in active (lectin) forms. The bonding process has a
certain probability $P\left( t\right) $. The simplest choice for the binding
probability in unit time is 
\begin{equation}
P\left( t\right) =\frac{\left[ W_{0}-N_{bound}\left( t\right) \right] }{W_{0}%
}\;\;,  \tag{2}
\end{equation}
$W_{0}-N_{bound}\left( t\right) $ is the number of the available receptors
(surface proteins) at the moment of time $t$. Consequently, the time
variation of the number of surface proteins (receptors) where calcium ions
are bound (receptor occupation), $N_{bound}\left( t\right) ,$ is given by
the following equation 
\begin{equation}
\frac{d}{dt}N_{bound}\left( t\right) =n_{Ca^{2+}}V_{ar}P\left( t\right) \;\;,
\tag{3}
\end{equation}
where $V_{ar}$ is the volume of the autocrine region. For numerical
applications we may approximate the radial dimension of the autocrine region
by the characteristic Debye length $\left( \lambda _{D}\right) $. The Debye
length measures the size of the ionic cloud which surrounds the (charged)
yeast cell. Doing so, $V_{ar}$ results in 
\begin{equation}
V_{ar}\cong \frac{4\pi }{3}\left[ \left( R+\lambda _{D}\right)
^{3}-R^{3}\right] \;\;,  \tag{4}
\end{equation}
where $R$ is the radius of (spherical) cell. The size of the ionic cloud
surrounding the cell is undoubtedly related to the $pH$ value of the medium.
Any change of the $pH$ value does affect both surface charge and Debye
length $\left( \lambda _{D}\right) $. Implicitly, the equilibrium value of
the calcium concentration around the cell is changed.

To obtain the solution of eq. $\left( 1\right) $, which gives us the number
of activated surface proteins, ready for flocculation, we have to integrate
numerically the equations system composed from eqs. $\left( 1\right) $, $%
\left( 2\right) $ and $\left( 3\right) $. In the assumptions of the present
model, $W_{0}$ entering eq. $\left( 2\right) $ has a constant value.

Before embarking on other details, we see that the activation of the surface
proteins which, in turn, promotes flocculation, depends in the present
kinetic model on: the total number of surface proteins ($W_{0}$), the
average concentration of calcium ions ($n_{Ca^{2+}}$), the rate and/or
efficiency of the cell to use sugars and the degree of saturation of sugar
content in the external medium ($\alpha _{-1}$), and on the specificity of
the medium (temperature, viscosity, charge density on the cell membrane) by
the net value of the Debye length.

According to the cascade theory of flocculation the rate-limiting step of
the process is the combination of single cells to form doublets (Stradford,
1992). The doublets combine to form groups of four, and on to eight, $16,$ $%
32,$ $64,$ etc. (cascade theory). Subsequent collisions between pairs of
increasingly larger particles are energetically easier and are therefore not
rate-limiting.

The rate of the dimer formation is in a direct proportion with the
concentration of free cells $\left( c_{0}-c_{f}\right) $ (where $c_{0}$ is
the initial concentration of free cells and $c_{f}$ stands for the dimers at
the moment of time $t$), with the relative number of activated proteins in
unit time $\frac{1}{W_{0}}\frac{dW}{dt}$ and, indeed, depends on the
probability that free cells collide each other under thermal motion
conditions $\nu $. Therefore, this can be written as 
\begin{equation}
\frac{dc_{f}}{dt}\cong \frac{1}{W_{0}}\frac{dW}{dt}\left( c_{0}-c_{f}\right)
\nu \;\;\;.  \tag{5}
\end{equation}
The number of collisions which occur between molecules in a fluid system can
be calculated from a complex function (Chapman and Cowling, 1970) depending,
mainly, on the temperature $T$ and on the viscosity of the medium. This
resolves to 
\begin{equation}
\nu =aT^{\frac{1}{2}}\;\;,  \tag{6}
\end{equation}
where $a$ is a constant measured in appropriate units.

\section{Results}

The present model assumes that yeast flocculation is an kinetic process
depending on several effectors. The effectors may change both the initial
rate of flocculation and the net value of the content of remaining free
cells. Their influence will be examined in the following. The initial number
of free cells was set at $10^{9}$ per litter and any cell division was
disregarded at the present level of approximation. Also, we assumed that
each yeast cell has approximate $10^{6}$ surface proteins. The cell is
considered as having a spherical form with the radius equal to about $5\;\mu
m.$

\subsubsection{Cell efficiency to use sugar/nutrients}

We proceeded to integrate numerically $\left( 5\right) $ over a wide
interval of time, between $0$ and $100$ (arbitrary) time units. The calcium
content is maintained at a constant value $n_{Ca^{2+}}=3.1\;10^{21}$ ions
per unit volume. The Debye length was set at $\lambda _{D}=0.1\;\mu m$ and,
for simplicity, $\nu $ equal to unity. Generally, we observed that the
flocculation process proceeds from a high initial rate which progressively
declines until a steady state is reached where no further flocculation
occurs, leaving a small fraction of free cells. This general behavior agrees
with the experimental observations (see Stradford, 1992 and references
therein). Specifically, in Fig. 1 we may see the behavior of the free cells
concentration, $c_{0}-c_{f},$ for two different values of the $\alpha $
parameter. For $\alpha =0.9$, which means an almost ideal efficiency of cell
to use sugar/nutrients, we can see a high initial rate of flocculation. The
steady state is achieved relatively soon and is characterized by a small
fraction $\left( \frac{c_{0}-c_{f}}{c_{0}}\simeq 10^{-4}\right) $ of free
cells. (The last result is not evident from Fig. 1 .) Lowering the
efficiency of yeast cell to use sugar/nutrients to $\alpha =0.5,$ the value
of the remaining free cells fraction is drastically enhanced about three
orders of magnitude $\left( \frac{c_{0}-c_{f}}{c_{0}}\simeq 10^{-1}\right) .$
In the latter case, the initial rate of flocculation is dramatically
changed, the slope being modified with about $20\%$.

\subsubsection{Calcium content}

In the following, we will keep the efficiency parameter at a constant value $%
\left( \alpha =0.5\right) $. The concentration of the calcium ions is now
varied while, all the parameters in above remain at their previous values.
We increased the calcium content of three times, from $3.1\;10^{21}$ ions to 
$9.3\;10^{21}$ ions per unit volume. Looking at Fig. 2, we may observe that
the initial rate of flocculation is strongly influenced by the calcium
content, as we just expected. Supplying the calcium content at the initial
stage of evolution of the cell culture, the activation rate of the surface
proteins is sped up resulting in a more rapid flocculation process. The
steady state free cells fraction is almost the same for both calcium
contents.

The same effect, as above, can be achieved by changing the Debye length and,
implicitly, the volume of the autocrine region. (In the practicality the
Debye length should be related to the $pH$ value in the external medium.)

\section{Final remarks}

Although sporadic flocculation may appear even from an earlier stage of the
yeast culture development, it is actually initiated after the growth process
ceased. The quantity of free cells in solution decreases sharply after a
certain time has elapsed. Under brewing conditions, the initiation of this
process is triggered after the growth limitation proceeded by a limited
oxygen supply (i.e., oxygen saturation of the wort at the beginning of
fermentation), as it was shown recently (Straver {\it et. al}, 1993a). The $%
CO_{2}$ formation during the fermentation produces a natural agitation among
the suspended cells that is a causal factor in flocculation. Agitation
causes rapid and progressive flocculation but, if at any time, due to
various reasons, agitation ceased, flocculation stopped (Stradford, 1987).
In the present model, the above observation can be easy correlated with the
appropriate number of collisions $\nu ,$ which occur between cells in the
suspension. By increasing $\nu $ the rate of flocculation goes into higher
values.

On the other hand, the delay in the initiation of flocculence has been seen
as an expression of the fact that the synthesis of the lectin involved in
flocculation of brewer's yeast is not regulated during the growth process
(Straver {\it et. al}, 1993a). This observation indicates that the
regulation of flocculation might be controlled at the level of the
expression of cell-surface activation abilities. Characterization and
regulation of flocculation at this level of surface cell activation is a
real challenge in the development of industrially feasible methods for
manipulating yeast-cell components in order to control flocculence during
fermentation. The theoretical model of flocculation developed in the present
paper offers, at a certain extent, an useful tool for further investigations
on this line.

Moreover, a sum of other similar biological processes can be described by
using the present kinetic approach. For example, the adhesion of yeast cells
to carbon dioxide bubbles (flotation) in the wine-making technology or the
binding of cells to the matrix and to microorganisms that have already
adhered can be subject of the present approach.

\newpage

{\large REFERENCES}

Cantrell, D. A. and Smith, K. A., 1984, The Interleukin-2 T-Cell System: a
New Cell Growth Model, Science 224{\bf ,} 1312-1316.

Chapman, S. and Cowling T.G., 1970, The Mathematical Theory of Non-Uniform
Gases, 3rd ed. (Cambridge University Press) pp. 235-236.

Davis, R.H. and Hunt, T.P., 1986, Modeling and Measurement of Yeast
Flocculation, Biotechnology Progress 2, 91-97.

Despa, S.-I.., and Despa, F., 1997, Diffusion Model for Growth Factors-Cell
Receptors Interaction, BioSystems, 44 59-68.

Harris, J.O., 1959, Possible Mechanism of Yeast Flocculation, Journal of the
Institute of Brewing 65, 5-6.

Mandelbrot, B.B., 1990, Fractals - a Geometry of Nature, New Scientist 127,
38-43.

Miki, B.L.A., Poon, N.H., James, A.P. and Selegy, V.L., 1982, Possible
Mechanism for Flocculation Interactions Governed by the Gene FLO1 in {\it %
Saccharomyces cerevisiae, }J. Bacteriol. 150{\bf ,} 878-889.

Mill, P.J., 1964, The Nature of the Interactions between Flocculent Cells in
the Flocculation of {\it Saccharomyces cerevisiae}, Journal of General
Microbiology 35, 61-68.

Novick, P., Ferro, S. and Schekman, R., 1981, Order of Events in the Yeast
Secretory Pathway, Cell 25{\bf ,} 461-469.

Schaefer, D.V., 1989, Polymers, Fractals, and Ceramic Materials, Science
243, 1023-1027.

Stradford, M., 1992, Yeast Flocculation: A New Perspective, Adv. Microb.
Physiol. 33, 2-71.

Stratford, M. 1992a, Yeast Flocculation: Reconciliation of physiological and
Genetic Viewpoint, Yeast\ 8, 25-38.

Stratford, M. and Keenan, M.H., 1987, Yeast Flocculation: Kinetics and
Collision Theory, Yeast 3, 201-206.

Straver, M.H., Kijne, J.W. and Smith, G., 1993, Cause and Control of
Flocculation in Yeast, Trends in Biotechnology 11, 228-232.

Straver, M.H., Smit, G. and Kijne, J.W., 1993a, Determinants of Flocculence
of Brewer's Yeast During Fermentation in Wort, Yeast 9, 527-532.

\newpage

{\Large FIGURE CAPTIONS}

\bigskip

Fig. 1 - The influence of the efficiency of the yeast cell to use the
sugar/nutrients on the flocculation process.

\bigskip

Fig. 2 - The influence of the calcium content on the flocculation process.

\end{document}